\documentclass{article} %
\usepackage{iclr2027_conference,times}
\usepackage[T1]{fontenc}

\usepackage{amsmath,amssymb}
\usepackage{graphicx}
\usepackage{booktabs}
\usepackage{array}
\usepackage{float}
\usepackage{microtype}
\usepackage{balance}
\usepackage{hyperref}
\usepackage{url}
\usepackage{subcaption}
\usepackage{makecell}

\title{ToolSched: Resource-Efficient Tool Scheduling for Shared-Host LLM Agents}

\author{Shucheng Zou$^{1}$, Yuhao Gu$^{1}$, Long Chen$^{1}$, Zhongchun Zheng$^{1}$, BinZe Xie$^{2}$, Xianwei Zhang$^{1}$\\
$^{1}$SUN YAT-SEN UNIVERSITY (mail.sysu.edu.cn)\\
$^{2}$Independent (independent-researcher.org)
}

\iclrfinalcopy %
\begin{document}

\pagestyle{plain}
\maketitle
\thispagestyle{plain}
\begin{abstract}

Local LLM agents often work on shared hosts. Unlike multi-tenant cloud clusters, these hosts often have tightly bounded memory, CPU, and I/O resources. Concurrent agents frequently suffer from resource contention, leading to failures such as OOM and reduced resource efficiency. Considering that agents' resource demands are bursty and shell command delay is imperceptible to agents, there is a huge scheduling opportunity to arrange when and in what order commands are actually executed. Thus, we present ToolSched, a Shell MCP Server for shared local hosts that serves multiple agents and schedules their operations. ToolSched embeds a resource model into the MCP interface: agents estimate resource demands along with command requests. These hints, together with real-time host resource conditions, determine when commands are actually executed. Internally, ToolSched applies a scheduling algorithm to stagger commands, reducing contention and improving resource utilization. Prediction correction and checkpoint-based recovery mitigate estimation bias and resource errors. Our evaluation on Terminal Bench shows that, compared with unscheduled execution, ToolSched improves resource utilization by 148.4\% and end-to-end command throughput by 29.4\%, demonstrating its effectiveness and practicality.

\end{abstract}

\section{Introduction}\label{sec:introduction}

Large language models (LLMs) have evolved from text generators into multi-turn chat bots~\citep{openai2022chatgpt}, and are now entering the era of agents~\citep{wang2024surveyagents}. Unlike previous LLM services that merely generated text, agents connect the thinking ability of LLMs with real-world operations~\citep{yao2023react}: they call tools or execute commands to interact with environments, and iteratively revise their behavior to complete complex tasks. This capability enables agents to handle domains such as software engineering~\citep{yang2024sweagent,jimenez2024swebench,wang2025openhands}, data analysis~\citep{chan2025mlebench,hong2025datainterpreter}, and scientific experimentation~\citep{boiko2023autonomous}, where success depends on continuous interaction with stateful execution environments. Nowadays, most ChatGPT-like chat applications already possess some agent capabilities~\citep{openai2025chatgptagent}. They run in the cloud, co-located with inference services, respond to remote users' dialogue requests, and call predefined tools during responses to improve their answers.

\begin{figure*}[t]
    \centering
    \includegraphics[width=\textwidth]{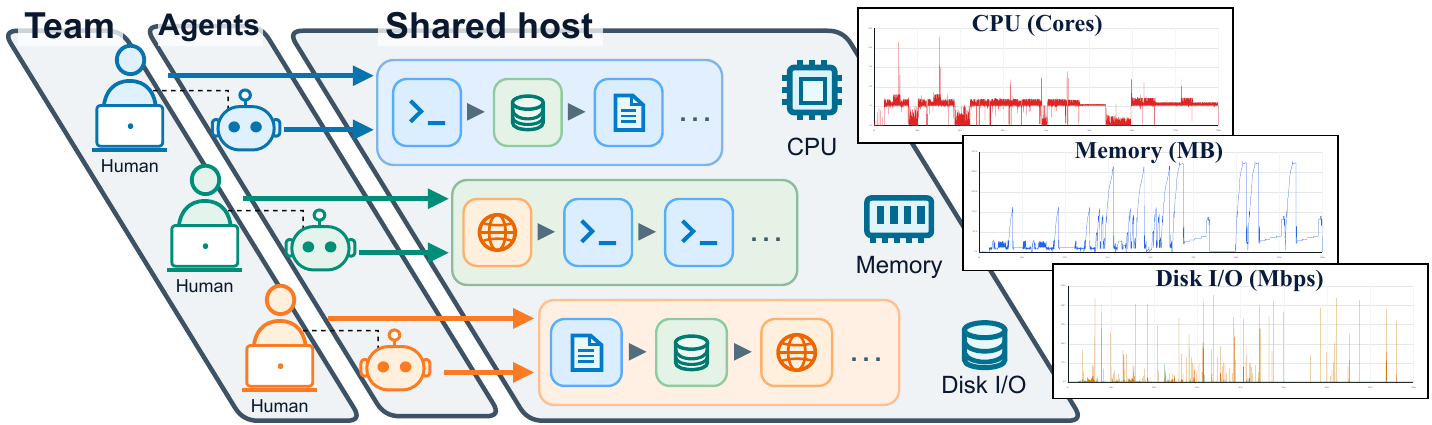}
    \caption{A group of team members may share a single host machine, along with their agents. The right part of the figure presents typical observations during a real task, where resource contention may arise due to highly dynamic resource demands.}
    \label{fig:introduction}
\end{figure*}

Another class of agents runs as standalone applications on local hosts, acting as a bridge between LLM inference services and real-world operational environments: they request external LLM inference services and translate the output tokens into concrete actions like editing files or executing shell commands on behalf of users, such as OpenClaw~\citep{openclawrepo}, Hermes~\citep{hermesagentrepo}, etc. In such scenarios, it is common to see multiple agents work on the same host. As shown in Figure~\ref{fig:introduction}, a workgroup or a small team may share a server used jointly by all members and their agents. Unlike multi-tenant cloud clusters, the computational resources, such as memory, CPU, and disk I/O, are highly constrained. Contention among multiple agents often leads to execution failures due to resource insufficiency, i.e., out-of-memory (OOM) errors, as observed by \citet{zheng2026agentcgroup}.

Coordination among agents is demanded to alleviate this resource contention problem. Figure~\ref{fig:introduction} demonstrates that the resource demands during a single agent task can fluctuate dramatically: memory, CPU, and I/O demands may differ by an order of magnitude or more across stages. On the other hand, since an LLM's inference output is only related to its input, from the agent's perspective, delaying the execution of a shell command is semantically indistinguishable from the command itself taking longer to execute. This presents a significant scheduling opportunity: by appropriately arranging the execution order and time of shell commands, we can coordinate resource usage among multiple agents without affecting the agents' final behavior, thus reducing contention, improving resource utilization, and ultimately increasing overall performance.

Driven by this idea, we propose ToolSched, a Shell MCP~\citep{mcptools2025} Server running on local shared hosts. ToolSched can serve multiple agents simultaneously and manage their associated Linux containers used for executing shell commands. Unlike previous Shell MCP Servers, ToolSched embeds a resource model into the MCP interface: it requires the agent to estimate the resource requirements when issuing an operation request. These hints, together with the host's real-time resource conditions, are used to decide when to actually execute the shell command. Internally, ToolSched applies a scheduling algorithm to achieve staggered execution, thereby reducing contention and improving host resource utilization. To mitigate estimation bias and overload challenges, ToolSched also introduces prediction correction and checkpoint-based recovery mechanisms to improve scheduling accuracy and system reliability.

In conclusion, the contributions of this paper are as follows:
\begin{enumerate}

    \item We propose that it is possible to apply proactive shell command scheduling to alleviate resource contention among LLM agents operating on a shared host, without changing the semantics of commands as visible to agents.

    \item We introduce ToolSched, a prototype Shell MCP Server that can coordinate multiple agents and, through real-time resource-aware scheduling, achieve accurate and reliable staggered execution of shell commands.

    \item Our evaluation on Terminal Bench~\citep{merrill2026terminalbench} shows that ToolSched can improve resource utilization by 148.4\% and command throughput by 29.4\%, compared with unscheduled execution, proving its effectiveness and practicability.

\end{enumerate}
\section{Background}\label{sec:preliminaries}

\textbf{Local LLM Agents with Shared Hosts.} In this paper, we focus on agents deployed on local machines, such as OpenClaw~\citep{openclawrepo}. These agents serve as a bridge between LLM inference services and local environments, translating LLM outputs into operations like executing shell commands. Just as human members of a team may share a physical server, agents may also operate on the same machine. In such settings, Linux containers~\citep{podmandocs} are commonly used to prevent different tasks from interfering with one another. As a lightweight isolation mechanism, containers can create independent operating environments while allowing limited sharing of files and other resources among them~\citep{soltesz2007containers}, making them well suited to the needs of agents.

\textbf{Resource Contention.} Since these agents operate on the same host, resource contention inevitably arises. A complete task workflow typically involves executing many shell commands, whose resource demands can differ dramatically~\citep{zheng2026agentcgroup}—for example, moving a file with \texttt{mv} versus compiling a program with \texttt{gcc}. Because the host's total resources are limited, when multiple agents operate simultaneously, some commands may fail or run inefficiently due to resource insufficiency~\citep{delimitrou2013paragon,kubernetesresources}. Human users usually sense such contention and always can coordinate usage through communication. However, agents generally do not possess such inter-communication ability—an agent running inside a container is, by default, unaware of the existence of agents in other containers. When resource-related failures occur, without awareness of the root cause, the agent may fall into continuous retries, further intensifying resource contention and even causing large-scale workflow blockage.

\textbf{Tool-Call Schedulability.} On the other hand, there is substantial scheduling slack among these operations. The fundamental reason is that LLM inference is essentially predicting subsequent output based on input context, and the result of a tool call is merely fed back into the model as new context~\citep{schick2023toolformer}; thus, it is almost entirely insensitive to timing. Taking OpenAI's \texttt{/chat/completions} API as an example, after the model generates a tool call, the agent is only required to include a message with the \texttt{tool} role in its next request~\citep{openaifunctioncalling}, while no requirement is imposed on when the next round of conversation is requested. The same principle applies to Shell MCP Servers, which are commonly used by agent applications (such as OpenClaw) to operate remote machines. Therefore, an MCP server can choose to delay command execution or reorder received operations without producing any visible semantic or behavioral change for the agent, which is often indistinguishable from the command itself taking somewhat longer to execute or the network communication being somewhat slower, and therefore does not alter the agent’s workflow or final outcome.

\section{Design and Implementation}\label{sec:method}

\subsection{Overview}\label{sec:overview}

\begin{figure*}[!h]
    \centering
    \includegraphics[width=0.98\textwidth]{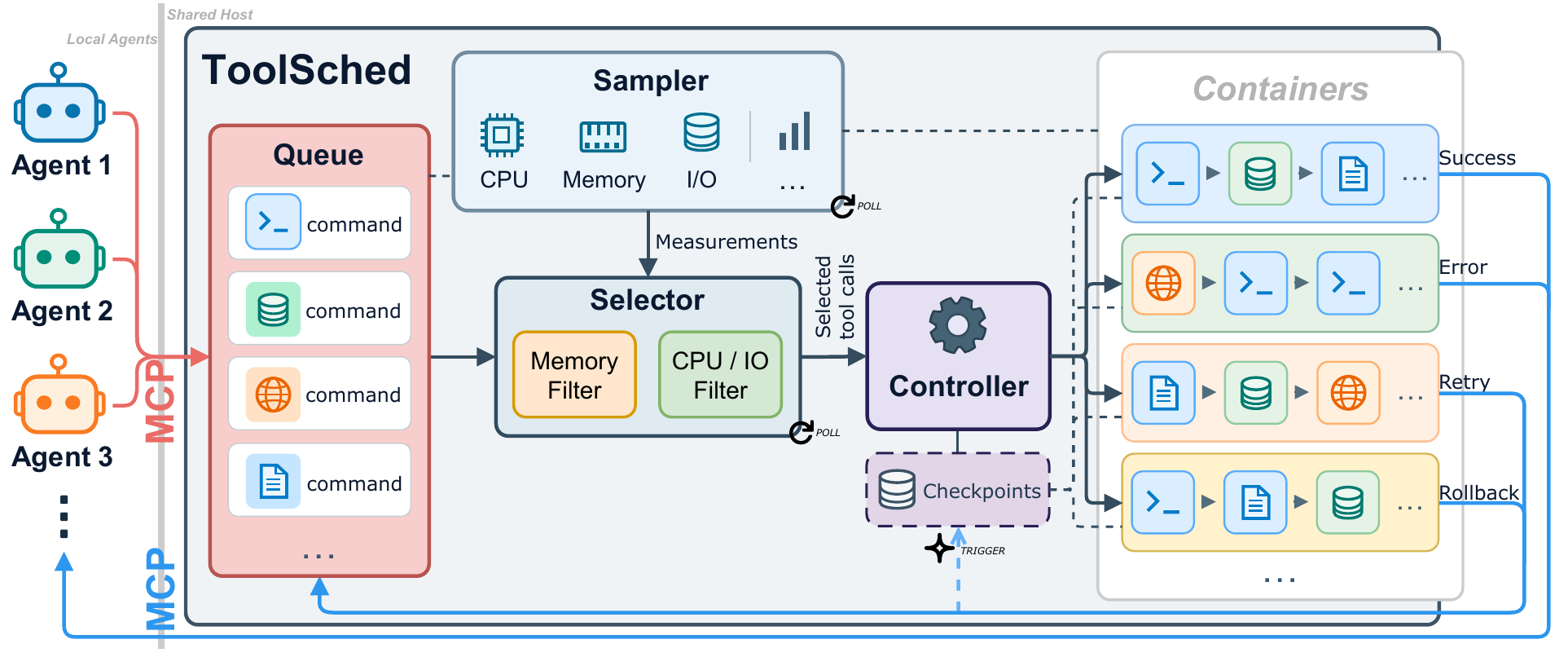}
    \caption{The architecture of ToolSched. ToolSched works on the shared host, interacts with agents via the MCP protocol and proxies their operations to their respective container environments.}
    \label{fig:overview}
\end{figure*}

ToolSched is a shell MCP server running on the host shared among agents. Just like other shell MCP servers~\citep{mcptools2025}, it allows agents to request shell commands and forwards them to the actual execution environment (specifically, containers). However, unlike existing works, ToolSched requires agents to declare the resources needed, including memory, CPU, and disk I/O, along with the shell commands in the MCP, which form the basis for ToolSched to schedule commands. In addition, since containers may have processes that need to run in the background, agents are also required to declare whether the command is idempotent and the recovery commands to set up background processes. 

Figure~\ref{fig:overview} shows the internal architecture of ToolSched, which consists of the Queue, Selector, Sampler, and Controller. When agents submit shell commands to ToolSched, these commands are not executed immediately; instead, they are pushed into the Queue for processing. The Selector then polls the entire queue and selects eligible commands to emit to the Controller for execution. The Sampler is in charge of system resource monitoring and provides real-time information for making reasonable selections. The Controller is responsible for dispatching commands to the corresponding container, supervising their execution and handling resource errors.

ToolSched classifies the execution result of a command into four categories:
\begin{enumerate}
    \item \textbf{Success:} The command runs successfully without any error, and then it flows out of ToolSched.
    \item \textbf{Error:} The command encounters an error during execution; it also flows out of ToolSched, but with the collected relevant error information.
    \item \textbf{Retry:} An idempotent command is aborted due to a resource error; in this case, the command is pushed back into the queue and re-executed once resources become available.
    \item \textbf{Rollback:} A non-idempotent command is aborted due to a resource error; in this case, the corresponding container is first rolled back by the Controller, and then the command is pushed back into the queue for re-execution.
\end{enumerate}

Only when a command flows out does ToolSched provide feedback to the agents, i.e., by generating MCP responses. For the latter two cases, the agent is completely unaware that the command has been re-enqueued. If the command eventually succeeds, everything is fine; if it ultimately fails, all error information within is collected and returned to the agents. It is believed that the agent is intelligent enough to handle these errors.

\subsection{Selector and Scheduling Algorithm}\label{sec:d1}

The Selector operates in a polling manner with a polling interval of $T_{sel}$. In each poll, it traverses the entire task queue, selects the currently eligible tasks, and emits them to the Controller for execution. %
Its goal is to maximize system resource utilization and concurrency. So we divide system resources into two categories: hard resources and soft resources. Hard resources currently include only memory, whose runtime constraints must be satisfied~\citep{kubernetesresources}; otherwise, the command cannot run. Soft resources, such as CPU cores and disk I/O, benefit from time-division multiplexing~\citep{linuxcgroupv2}. Even if their total demand exceeds the physical hardware limit, they do not prevent execution; they merely slow it down.

\begin{figure*}[!h]
    \centering
    \includegraphics[width=\textwidth]{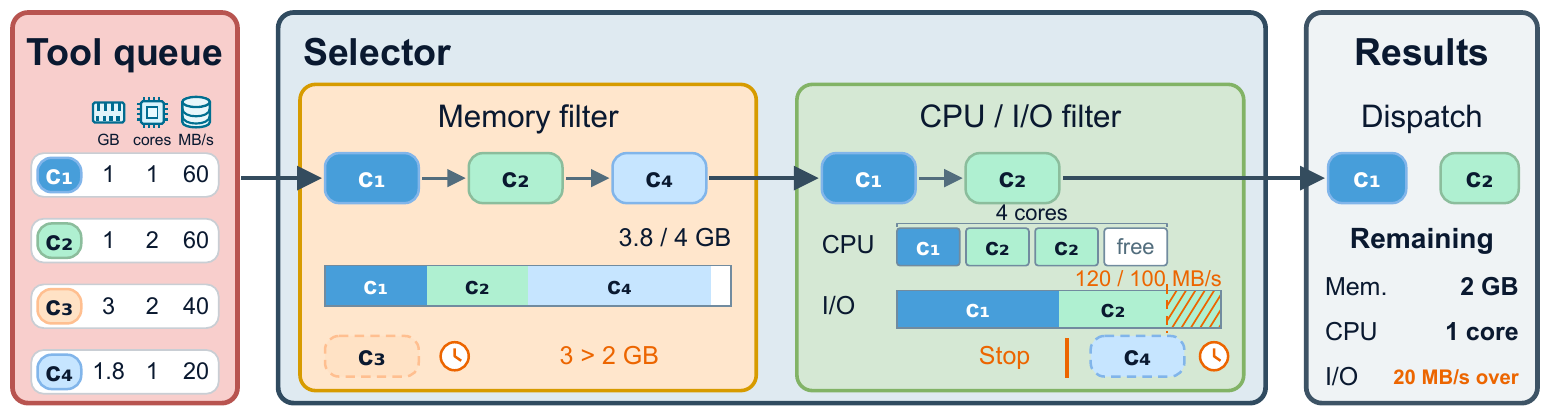}
    \caption{The Selector consists of two filters for the hard resource (memory) and soft resources. At each poll, all eligible commands are emitted to the Controller for execution.}
    \label{fig:scheduling}
\end{figure*}

For these two types of resources, the Selector adopts different scheduling strategies. As shown in Figure~\ref{fig:scheduling}, for hard resources, the Selector first filters out runnable tasks from the queue. Specifically, the filter first computes the current remaining amount of hard resources (i.e., total amount minus reserved amount). Then it traverses the queue from front to back. If the remaining amount can satisfy the demand $D_i$ of the task being visited, the task is considered eligible, and the remaining amount is reduced accordingly. When the traversal finishes, a list of eligible commands is generated.

Subsequently, the Selector traverses the eligible list from front to back and emits commands based on soft resource demands. To do so, the Selector first calls the Sampler to provide the current remaining amount $R(t_n)$ for each resource. Then it traverses the eligible list from front to back and similarly deducts the remaining amounts. The difference here is that the traversal stops when any resource's remaining amount becomes negative, i.e., the total amount of soft resources is allowed to exceed 100\% once. This design prevents large tasks from being blocked in the queue for a long time due to insufficient resources. Even if their demand exceeds the total system resources, they still have a chance to be emitted and executed.

\subsection{Sampler and Prediction Correction}\label{sec:d2}

The Sampler also operates in a polling manner. It interfaces with the operating system (container cgroups)~\citep{linuxcgroupv2} and samples the actual resource usage of each running command. The Sampler mainly serves two purposes: first, it computes the remaining amount for each resource; second, it corrects the agent's resource demand declarations based on actual observations, thereby mitigating the problem of inaccurate resource demand estimation by the agent.

The remaining amount of the hard resource, memory, at time $t$ is calculated as follows:
\begin{equation}
R_{\mathrm{mem}}(t)
=
C_{\mathrm{mem}}
-
\sum_{i\in\mathcal{E}(t)} D'_{i,\mathrm{mem}}.
\end{equation}
The remaining amount of each soft resource $r$ at time $t$ is calculated as follows:
\begin{equation}
R_r(t)
=
C_r
-
\frac{1}{T_{sam}}
\int_{t_n-T_{sam}}^{t_n} U_r(x)\,\mathrm{d}x
-
\sum_{i\in\mathcal{A}_n(t)} D'_{i,r}.
\end{equation}
Here, $C_r$ is the resource capacity, $D'_{i,r}$ is the corrected demand of command $i$, and $U_r(x)$ is the observed aggregate usage. $\mathcal{E}(t)$ contains all currently running commands, while $\mathcal{A}_n(t)$ contains commands admitted in the current window whose reservations have not been released. $T_{sam}=10$ seconds is the window length, and $t_n$ is its starting time. The integral is computed by discrete accumulation in practice.

For memory, peak demand is reserved throughout command execution to reduce the risk of exceeding capacity. For CPU and disk I/O, the previous window's measured average and newly admitted demands jointly stagger heavy commands within a window. At each window boundary, the measured baseline is updated, and estimated demands are accumulated only for commands admitted in the new window. If demand is overestimated, the lower actual usage measured in the next window allows more commands to be admitted. If demand is underestimated, the higher measured usage reduces the budget for new commands.

Prior work such as Leopard uses historical CPU and memory usage to estimate invocation resource load for admission control~\citep{cao2025leopard}. Inspired by this, the Sampler records the observed resource peak $P$ and compares it with the resource demand $D$ declared by the agent, thereby correcting future agent resource demand estimates. Specifically, the $i$-th resource demand $D_{i}$ is corrected to $D'_{i}$ using the previous $k$ command histories:

\begin{equation}
D'_{i} = D_{i} * \mathrm{GM} \left( \frac{P_{i-1}}{D_{i-1}}, \frac{P_{i-2}}{D_{i-2}}, ..., \frac{P_{i-k}}{D_{i-k}} \right)
\end{equation}

where $\mathrm{GM}$ denotes the geometric mean, and $k = 256$ in this paper. Finally, $R(t_n)$ and $D'_{i}$ are provided to the Selector as references for selection decisions. %

\subsection{Controller and Container Recovery}\label{sec:d3}

The Controller supervises the execution of commands and responds to possible resource errors. To handle such errors, the Controller must be able to roll back and recover the container state. We choose to perform this rollback and recovery at the granularity of a shell command: before each command execution, the Controller creates a checkpoint for the container. Specifically, the container state consists of two parts: the runtime state, i.e., the memory of all running processes, and the file system state, i.e., the container's \texttt{rootfs}. Unlike previous approaches, we do not checkpoint the memory (i.e., hot checkpoints such as CRIU~\citep{podmancheckpoint}); instead, we only snapshot the \texttt{rootfs}. The reason is that the memory is bulky and its checkpointing is extremely expensive~\citep{wu2026crab}. Meanwhile, the agent has error-handling capabilities and can actively restore the runtime state via the recovery commands.

With rollback and recovery capabilities in place, when a container experiences a resource error such as OOM, the Controller can reclaim some resources by releasing certain containers. It should be noted that the container experiencing the resource error is not necessarily the one causing the problem; it could be another container that exceeds its resource usage and squeezes others. Therefore, simply stopping the error container may not be sufficient. The Controller also needs to find the container that actually caused the problem and terminate it. 

Early Linux OOM heuristics combine memory consumption with CPU time
and elapsed runtime to favor memory-heavy, short-lived processes for
termination~\citep{gorman_oom}. Inspired by this approach, the Controller
considers both excess resource demand and the execution time of the
current command. Specifically, at the moment $t$ when the resource shortage occurs, for each hard resource we use the following formula to select the container with the largest $V(t)$ and release it:

\begin{equation}
V(t) = \frac{U(t) - D_i}{T_i}
\end{equation}

where $U(t)$ is the actual usage for the current command, $D_i$ is the declared demand, and $T_i$ is the time the container has been running for this command. For memory, $U(t)$ and $D_i$ are both measured as additional demand above the container's usage immediately before the command. After the command is aborted, the Controller re-enqueues the aborted command and sets its resource demand to $\max(U(t), 2 D_i)$. If the command is non-idempotent (declared by the agent in the MCP request), the Controller rolls back the container state to the previous checkpoint before re-enqueuing it.

\section{Evaluation}\label{sec:evaluation}

\subsection{Methodology}\label{sec:methodology}

We evaluate ToolSched from three aspects: 1) Its effect on resource utilization, and the improvement in performance brought by the increased utilization (\S\ref{sec:exp1}). 2) Its Sampler's prediction accuracy (\S\ref{sec:exp2}). 3) The overhead of the Controller's recovery mechanism (\S\ref{sec:exp3}).

\subsubsection{Testbed and Workload}\label{sec:testbed-workload}

\begin{table*}[htbp]
    \centering
    \caption{Testbed and Workload Configurations.}
    \begin{subtable}[b]{0.7\linewidth}
        \centering
        \caption{Testbed}
        \label{tab:testbed}
        \begin{tabular}{l|l|l|l|l}
            \toprule
            \textbf{CPU} & \textbf{Memory} & \textbf{Storage} & \textbf{LLM} & \textbf{Agent} \\
            \midrule
            \makecell[l]{Intel i7-10700 \\ 2.9GHz 8C16T}
                & 4GB
                & 256GB
                & GLM-5.2
                & \makecell[l]{Terminal \\ Bench Agent}
            \\
            \bottomrule
        \end{tabular}
    \end{subtable}
    \hfill
    \begin{subtable}[b]{0.28\linewidth}
        \centering
        \caption{Workload}
        \label{tab:workload}
        \setlength{\tabcolsep}{5pt}
        \begin{tabular}{l|l}
            \toprule
            \textbf{\makecell[l]{Suite}}
                & \textbf{\makecell[l]{Avail. / \\ Total Tasks}} \\
            \midrule
            \makecell[l]{Terminal \\ Bench 2.0} & 75 / 89 \\
            \bottomrule
        \end{tabular}
    \end{subtable}
    \label{tab:overall}
\end{table*}

\looseness=-1

Tables~\ref{tab:testbed} and~\ref{tab:workload} show the testbed and workload configurations used in our evaluation. Since the experiments aim to evaluate ToolSched rather than the capabilities of the LLM or the agents, we chose an older and easier version (2.0) of Terminal Bench~\citep{merrill2026terminalbench}. Even so, out of all 89 tasks, the LLM used still failed to complete 14 tasks within our time limit. Therefore, only the remaining 75 tasks were used. To support experiment replay and enhance reproducibility, we fully recorded the interaction between the agent and ToolSched for each task during the experiment.

\par

\subsubsection{Metrics and Schemes}\label{sec:metrics}

\looseness=-1

Our main concerns are resource utilization and the resulting performance improvement (task completion time). In addition, further evaluations of the Sampler and Controller focus on accuracy and time overhead, respectively, as described in Sections~\ref{sec:exp2} and~\ref{sec:exp3}.
We compare ToolSched (\texttt{toolsched}) with two other schemes: 1) \texttt{no-sched}: no scheduling at all, directly running all tasks simultaneously; 2) \texttt{reserve}: statically allocating the peak resource demand of each task, similar to practices in cloud clusters. 

\par

\begin{figure}[H]
    \centering
    \includegraphics[width=1\textwidth]{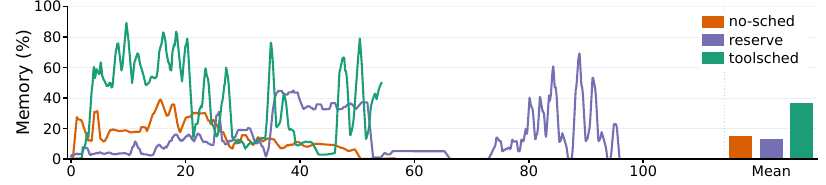}
    \includegraphics[width=1\textwidth]{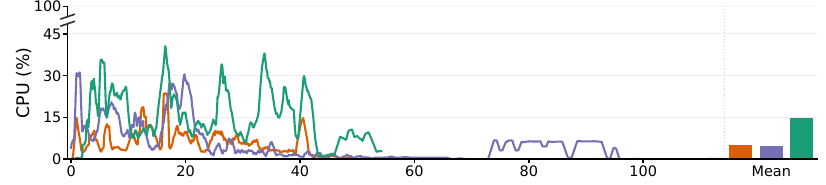}
    \includegraphics[width=1\textwidth]{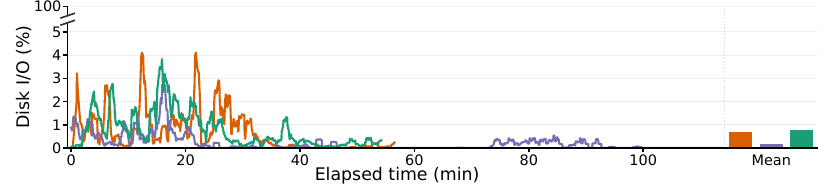}
    \caption{Observed resource usage throughout the entire experiment for the three schemes. It can be seen that \texttt{toolsched} achieves higher resource utilization for memory and CPU, and therefore completes all tasks the fastest. Since the I/O load of Terminal Bench is low and does not constitute a bottleneck, no significant difference is visible in the figure.}
    \label{fig:utilization}
\end{figure}

\begin{table}[H]
    \caption{Pass rate and performance results. Due to resource contention, \texttt{no-sched} fails on 11 tasks in Terminal Bench. By contrast, \texttt{toolsched} can complete all the tasks in less time.}
    \label{tab:performance}
    \centering
    \begin{tabular}{l|rrr}
        \toprule
        \textbf{Scheme}
            & \textbf{Pass (Succ. / Total)}
            & \textbf{Time (s)}
            & \textbf{Avg. Cmd./s}
        \\
        \midrule
        \texttt{no-sched}
            & 85.33\% (64/75)
            & 3,392
            & 0.462
        \\
        \texttt{reserve}
            & 100\% (75/75)
            & 6,837
            & 0.285
        \\
        \texttt{toolsched}
            & 100\% (75/75)
            & \textbf{3,254}
            & \textbf{0.598}
        \\
        \bottomrule
    \end{tabular}
\end{table}

\subsection{Utilization and Performance}\label{sec:exp1}

Figure~\ref{fig:utilization} depicts the resource usage observed for the three schemes throughout the experiment. Due to the resource contention problem, \texttt{no-sched} fails on some tasks, whose resource consumption has already been subtracted in Figure~\ref{fig:utilization}. The task pass rate, together with the total runtime and average commands per second, is reported in Table~\ref{tab:performance}.

From Figure~\ref{fig:utilization}, we can see that compared with \texttt{no-sched}, \texttt{toolsched} achieves a substantial improvement in memory and CPU utilization (mean memory utilization: 15.10\%, 13.16\%, and 36.67\%; mean CPU utilization: 5.24\%, 4.59\%, and 14.76\%, for \texttt{no-sched}, \texttt{reserve}, and \texttt{toolsched}, respectively). This shows that ToolSched achieves effective scheduling coordination, maximizing command concurrency while satisfying resource demands. The three schemes show no significant difference in I/O resource usage, because Terminal Bench has low I/O demand and is far from the testbed bottleneck.

In terms of overall performance, \texttt{toolsched} has a runtime similar to that of \texttt{no-sched} (3254~s vs. 3392~s in Table~\ref{tab:performance}). However, it must be noted that since \texttt{no-sched} failed on some tasks, \texttt{toolsched} actually completed more tasks and therefore achieved higher average shell command throughput (0.598 vs. 0.462). On the other hand, due to its overly conservative static resource allocation, \texttt{reserve} took 1.1x longer time than \texttt{toolsched} to complete all tasks, which proves the great space and opportunity for scheduling optimization at the sub-task command granularity.

\subsection{Prediction Accuracy}\label{sec:exp2}

\begin{figure}[H]
    \centering
    \includegraphics[width=1\textwidth]{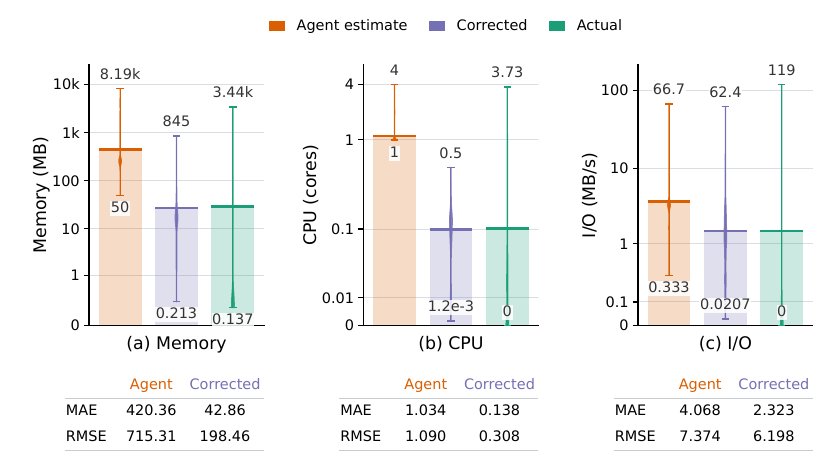}
    \caption{Comparison of agent estimates, corrected estimates, and measured resource usage across 1,573 tool calls. Bars show arithmetic means, narrow violins show distributions, and vertical lines span the observed ranges. The axes use a log1p transform, with ticks labeled in the original units. The tables report per-call mean absolute error (MAE) and root mean squared error (RMSE) in the corresponding resource units.}
    \label{fig:accuracy}
\end{figure}

To evaluate ToolSched's Sampler and its prediction correction mechanism designed in \S\ref{sec:d2}, we compare the observed real resource usage with ToolSched's corrected results, along with the agent's original estimates, as shown in Figure~\ref{fig:accuracy}. The figure intuitively shows the correction effect of ToolSched---it follows the overall trend of the estimates but is closer to the actual usage.

Further data analysis shows that the mean absolute error (MAE) between the agent's original estimates and actual usage is 420.36~MB (Memory), 1.034~cores (CPU), and 4.068~MB/s (I/O). After calibration, the MAE decreases to 42.86~MB (Memory), 0.138~cores (CPU), and 2.323~MB/s (I/O). Thus, the calibrated estimate provides a substantially more accurate signal for command scheduling than the agent's original estimation, which proves the necessity of the Sampler.

\subsection{Recovery Cost}\label{sec:exp3}

To evaluate the cost of the Controller's container recovery mechanism, we further measure the time overhead of container checkpoint/rollback and its occurrence frequency in the actual evaluation. The results are given in Table~\ref{tab:checkpoint-recovery}. On average, creating a checkpoint only slows down command execution by 54~ms, which is very small compared to the hundreds- or thousands-of-seconds task-completion time in the experiment.

Moreover, since the rollback rate during the entire task process is very low (0.46\%, 9/1946), the overhead introduced by container rollback is almost negligible. Overall, the introduction of the container recovery mechanism incurs only a very small overhead but solves the critical and challenging resource overload problem, serving as a stability and reliability fallback in ToolSched.

\begin{table}[!h]
    \centering
    \caption{Checkpoint and recovery overheads.}
    \label{tab:checkpoint-recovery}
    \begin{tabular*}{0.9\linewidth}{l|rrrrr}
        \toprule
        \textbf{Metric} & \textbf{Mean} & \textbf{P50} & \textbf{P95} & \textbf{Max} & \textbf{Frequency} \\
        \midrule
        Checkpoint latency (s)
            & 0.054 & 0.034 & 0.128 & 29.69 & 100\% (1946/1946) \\
        Checkpoint storage (MB)
            & 598.8 & 624.0 & 1011.6 & 1408.6 & --- \\
        Recovery latency (s)
            & 1.670 & 0.946 & 6.207 & 6.207 & 0.46\% (9/1946) \\
        \bottomrule
    \end{tabular*}
\end{table}

\section{Related Work}
\label{sec:related-work}

Agent tools have different resource demands across calls, making fixed resource allocations inefficient when multiple agents share a host.
Recent work addresses this variability at tool-call or action granularity.
One approach controls resource consumption during execution: AgentCgroup~\citep{zheng2026agentcgroup} places calls in separate cgroups and uses CPU scheduling and memory throttling or freezing to regulate running tools while preserving agent state.
Another approach adjusts resource allocation between actions: ARL-Tangram~\citep{xiao2026arltangram} allocates CPU cores when an action executes and releases them afterward.
It retains container memory to preserve environment state, a choice suited to its target clusters with relatively abundant memory.
These approaches make resource management responsive to individual calls, but differ in which resources they retain and when they intervene.
ToolSched focuses on coordinating command execution under the limited memory, CPU, and I/O capacity of a shared host.
It checks memory feasibility and schedules command start times according to resource availability, allowing selected commands to wait rather than compete immediately for resources.

Resource prediction helps determine which calls can execute concurrently.
Leopard~\citep{cao2025leopard} uses historical average resource usage to guide serverless invocation scheduling, while ALAP~\citep{sinha2025alap} predicts requirements from invocation inputs.
ToolSched similarly uses historical measurements, but corrects agent-declared demands for stateful shell execution rather than directly predicting demand from previous function invocations.
Such estimates can still be inaccurate, making recovery important when resource failures occur.
Existing agent recovery systems preserve filesystem and runtime state through checkpoints~\citep{dong2026deltabox,wu2026crab,zhou2026belayer}.
Crab reduces unnecessary checkpointing by observing state changes through eBPF, but still uses CRIU when process state needs to be captured.
ToolSched instead uses agent-provided recovery commands to reconstruct runtime state, so it snapshots only the rootfs without capturing process memory.
This avoids process-memory checkpointing and restoration, at the cost of relying on recovery commands to re-establish necessary background processes.
\section{Conclusion}\label{sec:conclusion}

We presented ToolSched, a Shell MCP server designed to coordinate multiple LLM agents running on a shared, resource-constrained host. Unlike conventional Shell MCP servers, ToolSched embeds a resource model into the MCP interface: agents provide resource estimates when issuing shell commands, and these hints, together with real-time host resource conditions, drive a scheduling algorithm that staggers command execution across agents. To handle estimation bias and overload, ToolSched further incorporates prediction correction and checkpoint-based recovery, improving both scheduling accuracy and system reliability. Our evaluation on Terminal Bench shows that, compared with unscheduled execution, ToolSched improves resource utilization by 142.9\%–181.4\% and command throughput by 29.4\%, demonstrating its effectiveness and practicality. These results suggest that proactive, resource-aware shell command scheduling is a promising direction for enabling efficient and reliable multi-agent execution on shared local hosts. Future work includes extending ToolSched to more diverse workloads, refining the prediction model, and exploring its integration with broader agent ecosystems.

\balance
\newpage

\section*{Reproducibility Statement}
The runtime semantics are specified in Sections~2--3, and the workload and evaluation protocol are reported in Section~4. The artifact will provide the code revision, configuration archive, workload manifest, image digests, execution commands, and event logs. The primary throughput metric and secondary latency metric use the post-preparation execution interval only; preparation, grading, and cleanup are recorded separately. The final submission will add repeated-run statistics.
\bibliography{main}
\bibliographystyle{iclr2027_conference}

\end{document}